\documentstyle[aps,12pt]{revtex}
\begin{document}
\title{Atomic spin squeezing in a $\Lambda $ system}
\author{A. Dantan$^{\dagger }$, M.\ Pinard$^{\dagger }$, V. Josse$^{\dagger }$, N.
Nayak$^{\ast }$, P.\ R. Berman$^{\ddagger }$}
\address{$^{\dagger }$Laboratoire Kastler Brossel, Universit\'{e} Pierre et Marie
Curie,\\
4 place Jussieu, F75252\ Paris Cedex 05 France\\
$^{\ast }$S. N. Bose National Centre for Basic Sciences, Calcutta, India\\
$^{\ddagger }$ Michigan Center for Theoretical Physics, FOCUS Center, and\\
Physics Department, University of Michigan, Ann Arbor, MI 48109-1120}
\maketitle

\begin{abstract}
Using a quantum theory for an ensemble of three-level atoms ($\Lambda $)
placed in an optical cavity and driven by electromagnetic fields, we show
that the long-lived spin associated with the ground state sublevels can be
squeezed. Two kinds of squeezing are obtained: self-spin squeezing, when the
input fields are coherent states and the atomic ensemble exhibits a large
non-linearity; squeezing transfer, when one of the incoming fields is
squeezed.
\end{abstract}


\section{\protect\bigskip Introduction}

In this paper, we present calculations performed on a model system
consisting of three-level atoms that interact with two fields. The atoms are
placed in an optical cavity which ensures a sizeable level of interaction,
in contrast with single pass schemes. Under such conditions, the validity of
quantum fluctuation calculations based on quantum Langevin equations is well
established. We will consider a set of three-level atoms in the so-called $%
\Lambda $ configuration, consisting of two ground state sublevels and one
excited state.

It is well-known that, in the case of large one-photon detuning (Raman
scheme), terms involving the ground - excited state coherence and excited
state population can be adiabatically eliminated. The three-level system is
then equivalent to an effective two-level system in which the spin
associated with the ground state sublevels may be squeezed. Because of its
simplicity the effective two-level system provides a good understanding of
the physical phenomena responsible for the creation of atomic squeezing and
allows one to carry out analytical calculations of optimal squeezing.

We study two regimes. In the first, the atoms interact with coherent light.
One can achieve atomic self-spin squeezing if the non-linearity of the
medium is sufficiently high. This atomic squeezing originates from the same
physical process that gives rise to the squeezing of the field exiting the
cavity, predicted theoretically in Refs.\cite{Reid,Lugiato,Hilico} and
observed experimentally a little later\cite{Lambrecht,Sinatra,Coudreau}. In
the early 1980's, it was conjectured that atomic spin squeezing appears as a
counterpart of squeezing of the electromagnetic field \cite{Walls,Carmichael}%
. However, the quantum noise reduction on atomic variables computed in
several papers can be obtained by a rotation of the atomic variables of a
two-level system interacting with a coherent field. As shown in \cite
{Kitagawa,Itano,Wineland}, in order to be relevant, spin squeezing should be
computed in the plane orthogonal to the direction of the mean spin. We have
performed full quantum calculations in the relevant basis. Here we
generalize the results obtained previously on two-level systems \cite{Vernac}
and three-level (V) \cite{Vernac3} systems to the $\Lambda $ system.

In the second regime, the atoms interact with a squeezed incoming field.
This scheme was proposed recently and experimentally tested in a single pass
configuration to produce spin squeezing between two excited levels \cite
{Polzik1,Polzik}. In \cite{Vernac2}, it was shown that the cavity
configuration should allow one to enhance the squeezing transfer between
two-level atoms and field. In both cases the transfer was limited by the
vacuum field fluctuations. We find here that the $\Lambda $ scheme allows
for a quasi-perfect squeezing transfer from the field to the atoms. Unlike
configurations in which the squeezing occurs between two excited states or
between a ground and excited state, the squeezing is created between two
long-lived ground states. Such a system offers significant advantages from
an experimental perspective.

\section{Equations for atomic fluctuations}

\subsection{Three-level atoms interacting with two fields}

We consider an ensemble of three-level atoms, the three levels being
labelled 1,2,3 in a $\Lambda $ configuration (Fig. 1). The atoms interact
with two light fields: an intense classical pump field $A_{1}$, having
frequency $\omega _{1}$ and a quantum probe field $A_{2}$, having frequency $%
\omega _{2}$. Field $A_{1}$ transverses the medium in a single pass and
drives the $1\rightarrow 3$ transition, while field $A_{2}$ is confined to
an optical cavity and drives the $2\rightarrow 3$ transition. The detunings
from atomic resonance are $\Delta _{1}=\omega _{31}-\omega _{1}$ for the
pump, and $\Delta _{2}=\omega _{32}-\omega _{2}$ for the probe. The cavity
resonance frequency closest to the probe frequency is $\omega _{c}$. The
cavity detuning for the probe field is $\Delta _{c}=$ $\omega _{2}-\omega
_{c}.$ The incoming quantum field that is coupled into the cavity is denoted
by $A_{2}^{in}$. The intensity of field $A_{1}$ is assumed to be much
greater than that of the quantum field.

The three-level system is described using nine collective operators for the $%
N$ atoms of the ensemble, the populations of levels $\left| 1\right\rangle ,$
$\left| 2\right\rangle $ et $\left| 3\right\rangle $, 
\begin{equation}
\Pi _{1}=\sum\limits_{i=1}^{N}\left| 1\right\rangle _{i}\left\langle
1\right| _{i},\;\;\Pi _{2}=\sum\limits_{i=1}^{N}\left| 2\right\rangle
_{i}\left\langle 2\right| _{i},\;\;\Pi _{3}=\sum\limits_{i=1}^{N}\left|
3\right\rangle _{i}\left\langle 3\right| _{i}\text{ ,}  \label{eqpop}
\end{equation}
the components of the optical dipoles in frames rotating at the laser
frequencies, 
\begin{equation}
P_{1}(t)=\sum\limits_{i=1}^{N}\left| 1\right\rangle _{i}\left\langle
3\right| _{i}e^{i\omega _{1}t},\;\;P_{2}(t)=\sum\limits_{i=1}^{N}\left|
2\right\rangle _{i}\left\langle 3\right| _{i}e^{i\omega _{2}t}\text{ }
\label{eqdip}
\end{equation}
(and their hermitian conjugates), and the operator associated with the
coherence between levels $\left| 1\right\rangle $ and $\left| 2\right\rangle 
$, 
\begin{equation}
P_{r}(t)=\sum\limits_{i=1}^{N}\left| 2\right\rangle _{i}\left\langle
1\right| _{i}e^{i(\omega _{2}-\omega _{1})t}  \label{eqdipr}
\end{equation}
(and its hermitian conjugate).

The Hamiltonian for the atom-field system is given by 
\[
H=\hbar \omega _{21}\Pi _{2}+\hbar \omega _{31}\Pi _{3}+\hbar \omega
_{2}a_{2}^{\dagger }a_{2}+\hbar g(P_{2}^{\dagger }A_{2}+A_{2}^{\dagger
}P_{2})+\hbar (\Omega _{1}^{\ast }P_{1}+\Omega _{1}P_{1}^{\dagger }), 
\]
where $g={\cal E}_{0}d_{23}/\hbar $ is a coupling constant, assumed real,
associated with the (quantum) cavity field, ${\cal E}_{0}=\sqrt{\hbar \omega
_{2}/2\epsilon _{0}{\cal S}c}$ is an amplitude appearing in the equation for
the cavity field operator $E_{2}={\cal E}_{0}\left( e^{-i\omega
_{2}t}A_{2}+e^{i\omega _{2}t}A_{2}^{\dagger }\right) $, $A_{2}$ and $%
A_{2}^{\dagger }$ are field creation and annihilation operators of having
commutator 
\[
\left[ A_{2}(t),A_{2}^{\dagger }(t^{\prime })\right] =e^{-\kappa \left|
t-t^{\prime }\right| }/\tau , 
\]
$\kappa $ is the cavity decay rate, $\tau =L/c,$ where $L$ is the cavity
length, ${\cal S}$ is the cross sectional area of the cavity field, $\Omega
_{1}=E_{1}d_{13}/\hbar $ is a coupling constant associated with the
classical pump field, $E_{1}(t)=\left( e^{-i\omega _{1}t}E_{1}+e^{i\omega
_{1}t}E_{1}^{\ast }\right) ,$ $d_{23}$ and $d_{13}$ are dipole moment matrix
elements. Note that $\Omega _{1}$ has units of frequency, while $g$ and $%
A_{2}$ have units of (frequency)$^{1/2}.$ The decay constants associated
with dipole operators $P_{1}$ and $P_{2}$ are taken to be equal and denoted
by $\gamma $. In order to account for the finite lifetime of the two ground
state sublevels 1 and 2, we include in the model another decay rate $\gamma
_{0}$, which is much smaller than $\gamma $. For example, $\gamma _{0}$ can
reflect transit-time broadening in the system; as such, it affects level 3
as well as levels 1 and 2. Typically, the atoms fall out of the interaction
area with the light beam in a time of the order of tens of milliseconds,
whereas $\gamma $ is of the order of the MHz for the excited state. We also
include incoherent pumping of levels 1 and 2 with rates $\Lambda _{1}$ and $%
\Lambda _{2}$, respectively, to allow for ground state atoms entering the
interaction volume. We will neglect any fluctuations in the total number of
atoms in the interaction volume and assume that $\Pi _{T}=\Pi _{1}+\Pi
_{2}+\Pi _{3}=\left( \Lambda _{1}+\Lambda _{2}\right) /\gamma _{0}=N.$

The time evolution of atomic polarizations and populations is obtained by
adding to the Heisenberg equations of motion, $i\hbar \dot{O}=[O,H],$ terms
corresponding to the Langevin forces associated with the decay of these
quantities,

\begin{equation}
\frac{d\Pi _{1}}{dt}=i\Omega _{1}^{\ast }P_{1}-i\Omega _{1}P_{1}^{\dagger
}+\gamma \Pi _{3}-\gamma _{0}\Pi _{1}+\Lambda _{1}+F_{\Pi _{1}}  \label{pi1}
\end{equation}
\begin{equation}
\frac{d\Pi _{2}}{dt}=igA_{2}^{\dagger }P_{2}-igA_{2}P_{2}^{\dagger }+\gamma
\Pi _{3}-\gamma _{0}\Pi _{2}+\Lambda _{2}+F_{\Pi _{2}}  \label{pi2}
\end{equation}
\begin{equation}
\frac{d\Pi _{3}}{dt}=-(i\Omega _{1}^{\ast }P_{1}-i\Omega _{1}P_{1}^{\dagger
})-(igA_{2}^{\dagger }P_{2}-igA_{2}P_{2}^{\dagger })-2\gamma \Pi _{3}-\gamma
_{0}\Pi _{3}+F_{\Pi _{3}}  \label{pi3}
\end{equation}
\begin{equation}
\frac{dP_{1}}{dt}=-(i\Delta _{1}+\gamma )P_{1}+i\Omega _{1}(\Pi _{1}-\Pi
_{3})+igA_{2}P_{r}^{\dagger }+F_{P_{1}}  \label{P1}
\end{equation}
\begin{equation}
\frac{dP_{2}}{dt}=-(i\Delta _{2}+\gamma )P_{2}+igA_{2}(\Pi _{2}-\Pi
_{3})+i\Omega _{1}P_{r}+F_{P_{2}}  \label{P2}
\end{equation}
\begin{equation}
\frac{dP_{r}}{dt}=-\left( \gamma _{0}-i\delta \right) P_{r}+i\Omega
_{1}^{\ast }P_{2}-igA_{2}P_{1}^{\dagger }+F_{P_{r}}  \label{pr}
\end{equation}
where 
\[
\delta =\Delta _{1}-\Delta _{2} 
\]
is the detuning between the ground state sublevels. ''In'' terms for $P_{r}$
resulting from spontaneous emission have not been included in Eq. (\ref{pr}%
), owing either to polarization selection rules or a hyperfine separation of
levels 1 and 2 that is much greater than $\gamma .$ To these equations must
be added the evolution equation for the cavity field.

Changes of the intracavity field result from the incoming field $A_{2}^{in}$%
, losses through the coupling mirror, and the source field associated with
the atomic polarization. The evolution equation for the intracavity field is 
\cite{Vernac} 
\begin{equation}
\frac{dA_{2}}{dt}=-(\kappa +i\Delta _{c})\text{ }A_{2}+\frac{ig}{\tau }P_{2}+%
\sqrt{\frac{2\kappa }{\tau }}A_{2}^{in}  \label{evA2}
\end{equation}
As is evident from this equation, fluctuations of the incoming field give
rise to a Langevin force for the cavity field.

Our aim is to obtain the fluctuations of the spin operators associated with
levels 1 and 2. For this, we need to determine the fluctuations of operators 
$P_{r},$ $P_{r}^{\dagger },$ $\Pi _{1}$ and $\Pi _{2}$. To obtain equations
for the fluctuations, we linearize equations (\ref{pi1}-\ref{pi3}), and (\ref
{P1}-\ref{pr}, \ref{evA2}) and their hermitian conjugates. We define a
10-dimensional vector $\left| \xi (t)\right] $ as 
\[
\left| \xi (t)\right] =|A_{2}(t),A_{2}^{\dagger }(t),P_{1}(t),P_{1}^{\dagger
}(t),P_{2}(t),P_{2}^{\dagger }(t),P_{r}(t),P_{r}^{\dagger
}(t),S_{z1}(t),S_{z2}(t)]^{T} 
\]
where 
\[
S_{z1}=\left( \Pi _{1}-\Pi _{3}\right) /2;\text{ \ \ \ \ \ }S_{z2}=\left(
\Pi _{2}-\Pi _{3}\right) /2 
\]
and then set 
\[
\left| \xi (t)\right] =\left| \left\langle \xi (t)\right\rangle \right]
+\left| \delta \xi (t)\right] 
\]
The steady state values, $\left\langle \xi (t)\right\rangle ,$ are obtained
from Eqs. (\ref{pi1}-\ref{pr}), (\ref{evA2}) in steady state, using the fact
that the mean value of the Langevin operators are zero; however, these
analytical solutions are of marginal use.

The fluctuation vector $\left| \delta \xi (t)\right] $ obeys an equation of
motion

\begin{equation}
\frac{d\left| \delta \xi (t)\right] }{dt}=-\left[ B\right] \left| \delta \xi
(t)\right] +\left| F_{\xi }(t)\right]  \label{evXi12}
\end{equation}
where $\left[ B\right] $ is the linearized evolution matrix of the
atom-field system and the column vector $\left| F_{\xi }(t)\right] $
contains the Langevin forces: 
\begin{eqnarray}
\left| F_{\xi }(t)\right] &=&|\sqrt{2\kappa /\tau }\delta A_{2}^{in}(t),%
\sqrt{2\kappa /\tau }\delta A_{2}^{in\dagger
}(t),F_{P_{1}}(t),F_{P_{1}^{\dagger }}(t),  \label{Fksi12} \\
&&F_{P_{2}}(t),F_{P_{2}^{\dagger }}(t),F_{P_{r}}(t),F_{P_{r}^{\dagger
}}(t),F_{S_{z1}}(t),F_{S_{z2}}(t)]^{T}  \nonumber
\end{eqnarray}

The correlation matrix $\left[ G(t)\right] $ of the fluctuations is: 
\begin{equation}
\left[ G(t)\right] =\left| \delta \xi (t)\right] \left[ \delta \xi (0)\right|
\end{equation}
The variances are obtained from the zero time correlation functions,
contained in the matrix $\left[ G(0)\right] \ $which satisfies:

\begin{equation}
\left[ B\right] \left[ G(0)\right] +\left[ G(0)\right] \left[ B\right]
^{\dagger }=\left[ D\right]  \label{eqG12(0)}
\end{equation}
where $\left[ D\right] $ is the correlation matrix of the Langevin forces : 
\begin{equation}
\left\langle \left| F_{\xi }(t)\right] \left[ F_{\xi }(t^{\prime })\right|
\right\rangle =\left[ D\right] \text{ }\delta (t-t^{\prime }).
\end{equation}
The sub-matrix $\left[ G_{c}(0)\right] \ $is defined as the $4\times 4$
lower diagonal block of $\left[ G(0)\right] $ and contains the variances of $%
P_{r},$ $P_{r}^{\dagger },$ $S_{z1},$ and $S_{z2}.$

\subsection{Spin squeezing}

\subsubsection{Definition}

In the same way as a squeezed state of the electromagnetic field is defined
by comparison with a coherent state, a squeezed spin state will be defined
as having fluctuations in one component lower than that of a coherent spin
state \cite{Wineland}. A coherent spin state for $N$ atoms is defined as a
product state of $N$ uncorrelated spins, in which the state vector for the $%
i $-th spin is an eigenstate the individual spin operator in the $(\theta
,\phi )$ direction: 
\begin{equation}
\sigma _{\theta ,\phi \ i}=\sigma _{xi}\sin \theta \cos \phi +\sigma
_{yi}\sin \theta \sin \phi +\sigma _{zi}\cos \theta
\end{equation}
having eigenvalue $+1/2$, with $\sigma _{x\text{ }i}=(\sigma _{+\ i}+\sigma
_{-\ i}^{\dagger })/2,\;\;\sigma _{y\text{ }i}=(\sigma _{+\ i}-\sigma _{-\
i}^{\dagger })/2i$. The coherent spin state can be obtained as a rotation
from the state having all spins aligned along the $z$ axis. This coherent
spin state is an eigenvalue of the collective spin operator $S_{\theta ,\phi
}=\sum_{i=1,N}\sigma _{\theta ,\phi \ i}$, with eigenvalue $S=N/2$ \cite
{Agarwal}$.$ It satisfies the minimum uncertainty relationship with
fluctuations equally distributed over any two orthogonal components normal
to the $(\theta ,\phi )$ direction, the variance of which is equal to $%
S/2=N/4$ . If one can squeeze the fluctuations of the total spin within the
plane orthogonal to the mean value, it will result in noise reduction in
spin measurements. The condition for spin squeezing is then \cite{Eberly} 
\begin{equation}
\smallskip \Delta S_{\alpha }\leq \ \left| \langle S_{Z}\rangle \right| /2
\end{equation}
where the axes have been rotated in such a way that the $Z$ axis is in the
direction of the mean spin and $\alpha $ represents a direction in the $X,Y$
plane. The quantity $\langle S_{Z}\rangle $ is then the mean value of the
spin and $S_{X}$ and $S_{Y}$ have zero mean values. Spin squeezing can occur
only for a spin ensemble with $N>1$ since it implies the emergence of
quantum correlations within the spin ensemble.

\subsubsection{Minimum variance calculation}

We calculate the variances $\Delta S_{X}$ and $\Delta S_{Y}$ of the spin
variables in the new reference frame. For this, we perform a rotation
defined by angles $\phi $ around the $z$ axis and $\theta $ around the $Y$
axis (defined by the first rotation) such that

\begin{equation}
\cos \theta =\langle S_{z}\rangle /\mid \langle \vec{S}\rangle \mid \;\;\cos
\phi =\langle S_{x}\rangle /S_{\phi }  \label{tetafi}
\end{equation}
with $\mid \langle \vec{S}\rangle \mid =\sqrt{\langle S_{x}\rangle
^{2}+\langle S_{y}\rangle ^{2}+\langle S_{z}\rangle ^{2}}$ and $S_{\phi }=%
\sqrt{\langle S_{x}\rangle ^{2}+\langle S_{y}\rangle ^{2}}$, where the mean
values are the solutions of the steady state equations. A spin component in
the $X,Y$ plane making an angle $\alpha $ with the $X$ axis has a variance
given by: 
\begin{equation}
\Delta S_{\alpha }=\cos ^{2}\alpha \ \Delta S_{X}+\sin ^{2}\alpha \ \Delta
S_{Y}+\sin 2\alpha \ {\rm Re}\left( \Delta S_{XY}\right)
\end{equation}
with $\Delta S_{\mu \nu }=<\delta S_{\mu }(0)\ \delta S_{\nu }(0)>\ (\mu
,\nu =X,Y)$. As a consequence the values $\alpha _{0}$ of $\alpha $ for the
spin components having maximal and minimal variances satisfy: 
\begin{equation}
\tan 2\alpha _{0}=\frac{2\ {\rm Re}\left( \Delta S_{XY}\right) }{\Delta
S_{X}-\Delta S_{Y}}  \label{alphazero}
\end{equation}
In order to investigate squeezing, we compare the minimal variance to $%
\left| \langle S_{Z}\rangle \right| /2$. The corresponding normalized
variance thus obtained is called $\Delta S_{\min }$. Spin squeezing is
achieved when $\Delta S_{\min }<1$.

For the three-level system, the spin fluctuations corresponding to the
atomic coherence between levels 1 and 2 are given by lower diagonal block $%
\left[ G_{c}(0)\right] $ of the correlation matrix. To go into the relevant
basis, one must first perform the transformation from $P_{r}$, $%
P_{r}^{\dagger }$, $S_{z1}$ , $S_{z2}$ to $S_{cx},S_{cy},S_{cz}$ with: 
\begin{equation}
S_{cx}=\frac{P_{r}+P_{r}^{\dagger }}{2},\text{ }S_{cy}=\frac{%
P_{r}-P_{r}^{\dagger }}{2i},\text{ }S_{cz}=\frac{\Pi _{2}-\Pi _{1}}{2}
\label{defSc}
\end{equation}
For this we define matrix $\left[ R_{1}\right] $ : 
\begin{equation}
\left[ R_{1}\right] =\frac{1}{2}\left( 
\begin{array}{cccc}
1 & 1 & 0 & 0 \\ 
-i & i & 0 & 0 \\ 
0 & 0 & -2 & 2
\end{array}
\right)  \label{matriceR}
\end{equation}
Then one goes into the basis $S_{cX},S_{cY}$, where the mean spin direction
is along OZ, by using the $\left[ R_{2}\right] $ matrix: 
\begin{equation}
\left[ R_{2}\right] =\left( 
\begin{array}{ccc}
\cos \theta \cos \varphi & \cos \theta \sin \varphi & -\sin \theta \\ 
-\sin \varphi & \cos \varphi & 0
\end{array}
\right)
\end{equation}
with angles $\theta $ et $\varphi $ given by Eq.(\ref{tetafi})

The atomic correlation matrix $\left[ G_{c}^{\bot }(0)\right] $ in the $%
S_{cX},S_{cY}$ basis is 
\begin{equation}
\left[ G_{c}^{\bot }(0)\right] =\left[ R_{2}\right] \left[ R_{1}\right] %
\left[ G_{c}(0)\right] \left[ R_{1}\right] _{hc}\text{ }\left[ R_{2}\right]
_{hc}
\end{equation}
The variances in the XY plane are then

\begin{equation}
\Delta S_{cX}=\left[ G_{c}^{\bot }(0)\right] _{1,1},\;\Delta S_{cY}=\left[
G_{c}^{\bot }(0)\right] _{2,2},\;\Delta S_{cXY}=\left[ G_{c}^{\bot }(0)%
\right] _{1,2}
\end{equation}
The minimal variance $\Delta S_{c~\min }$ corresponds to the angle $\alpha
_{0}$ given by Eq.(\ref{alphazero}).

\section{Adiabatic elimination in the Raman scheme}

\subsection{Effective two-level system}

In the limit that $\left| \Delta _{1,2}\right| \gg \gamma ,\left| \Omega
_{1}\right| $, one can adiabatically eliminate the ground-excited state
polarizations. In this limit, the excited state population is much less than
unity and can be neglected. We assume also that $\left| \delta \right|
=\left| \Delta _{1}-\Delta _{2}\right| \ll \left| \Delta \right| \equiv
(\Delta _{1}+\Delta _{2})/2$, since the values of $\left| \delta \right| $
to be considered are of order of $\gamma _{0}\ll \left| \Delta \right| $.
Performing this adiabatical elimination and neglecting terms of order $%
\gamma /\Delta ^{2}$, one obtains simplified equations for the sublevels
variables $S_{+}=P_{r}$, $S_{-}=P_{r}^{\dagger }$ , $S_{z}=(\Pi _{2}-\Pi
_{1})/2$ and $\Pi =\Pi _{2}+\Pi _{1}=N$:

\begin{equation}
\frac{dS_{+}}{dt}=-(\gamma _{0}-i\tilde{\delta})S_{+}+2i\tilde{g}%
A_{2}S_{z}+F_{S_{+}}  \label{S+}
\end{equation}
\begin{equation}
\frac{dS_{-}}{dt}=-(\gamma _{0}+i\tilde{\delta})S_{-}-2i\tilde{g}%
A_{2}^{\dagger }S_{z}+F_{S_{-}}  \label{S-}
\end{equation}
\begin{equation}
\frac{dS_{z}}{dt}=-\gamma _{0}S_{z}+\frac{\Lambda _{2}-\Lambda _{1}}{2}+i%
\tilde{g}(A_{2}^{\dagger }S_{+}-S_{-}A_{2})+F_{S_{z}}  \label{Sz}
\end{equation}
\begin{equation}
\frac{d\Pi }{dt}=-\gamma _{0}\Pi +\Lambda  \label{pi}
\end{equation}
where $\Lambda =\Lambda _{1}+\Lambda _{2}$, 
\[
\tilde{\delta}=\delta +\frac{\left| \Omega _{1}\right| ^{2}}{\Delta }-\frac{%
g^{2}A_{2}^{\dagger }A_{2}}{\Delta } 
\]
is the effective atomic detuning corrected for light-shifts, and 
\[
\tilde{g}=g\frac{\Omega _{1}}{\Delta } 
\]
is an effective coupling constant. The evolution equation (\ref{evA2}) for
the quantum field becomes

\begin{equation}
\frac{dA_{2}}{dt}=-(\kappa +i\Delta _{c})A_{2}+i\frac{\tilde{g}}{\tau }S_{+}+%
\sqrt{\frac{2\kappa }{\tau }}A_{2}^{in}  \label{A2}
\end{equation}
In principle, the atomic fluctuation operators contain terms related to
repopulation of the ground states via spontaneous emission; however these
contributions are down by $\left( \gamma /\Delta \right) ^{2}$ and are
neglected. As a consequence, the fluctuation operators are associated only
with the decay rate $\gamma _{0}$ and the repopulating processes; we shall
see that the diffusion matrix formed from these operators has a fairly
simple form.

The problem has been reduced to that of an ensemble of two-level atoms
interacting with one field in an optical cavity whose effective Rabi
frequency is $\Omega _{eff}=\tilde{g}\left\langle A_{2}\right\rangle =g\frac{%
\Omega _{1}}{\Delta }\left\langle A_{2}\right\rangle .$ We will use these
simplified equations for the spin squeezing calculations and give a physical
interpretation to our results. Note that, beside the incoherent pumping
terms, our two-level system differs slightly from a classical
''excited-ground state'' two-level system. The decay terms in (\ref{S+}-\ref
{S-}-\ref{Sz}) are all $\gamma _{0}$; the coherence relaxes at the same
rate\ as the populations. This remark, as simple as it may seem, critically
affects the steady states that are allowed by the system (see Sec. IV C).
Note also that the atomic noise scale is no longer $\gamma $, but $\gamma
_{0}$, hence much smaller. We thus expect the atomic noise not to destroy
the squeezing too much.\ 

\subsection{Steady state and bistability}

The steady state values are obtained solving Eq. (\ref{S+}-\ref{S-}-\ref{Sz}-%
\ref{A2}) in steady state, using the fact that the mean value of the
Langevin operators are zero. Explicitly, one finds 
\begin{equation}
\langle S_{+}\rangle =\frac{\Lambda _{2}-\Lambda _{1}}{\gamma _{0}}\beta _{2}%
\frac{i-\bar{\delta}}{1+\bar{\delta}^{2}+4I_{2}}  \label{statS+}
\end{equation}
\begin{equation}
\langle S_{-}\rangle =-\frac{\Lambda _{2}-\Lambda _{1}}{\gamma _{0}}\beta
_{2}\frac{i+\bar{\delta}}{1+\bar{\delta}^{2}+4I_{2}}  \label{statS-}
\end{equation}
\begin{equation}
\langle S_{z}\rangle =\frac{\Lambda _{2}-\Lambda _{1}}{2\gamma _{0}}\frac{1+%
\bar{\delta}^{2}}{1+\bar{\delta}^{2}+4I_{2}}  \label{statSz}
\end{equation}
with the dimensionless variables (in units of $\gamma _{0}$)

\begin{equation}
\bar{\delta}=\frac{\tilde{\delta}}{\gamma _{0}},\quad \beta _{2}=\frac{%
\tilde{g}\langle A_{2}\rangle }{\gamma _{0}},\quad I_{2}=\beta _{2}^{2}.
\end{equation}

For certain values of the parameters, the system exhibits bistability. From (%
\ref{A2}) in steady state, one derives the mean value of the intracavity
intensity 
\begin{equation}
I_{2}^{in}=I_{2}\left[ \left( 1+\frac{2\tilde{C}}{1+\bar{\delta}^{2}+4I_{2}}%
\right) ^{2}+\left( \delta _{c}+\frac{2\tilde{C}\bar{\delta}}{1+\bar{\delta}%
^{2}+4I_{2}}\right) ^{2}\right]  \label{bistability}
\end{equation}
where $\delta _{c}=\Delta _{c}/\kappa $ is the dimensionless cavity
detuning, and $\tilde{C}$ is an effective cooperativity or strong-coupling
parameter 
\begin{equation}
\tilde{C}=\frac{\tilde{g}^{2}}{2\kappa \tau \gamma _{0}}\frac{\Lambda
_{2}-\Lambda _{1}}{\gamma _{0}}
\end{equation}
We will limit our discussion to an incoherent pumping scheme in which $%
\Lambda _{1}=0$ and $\Lambda _{2}=N\gamma _{0},$ corresponding to absorption
for the probe field. We will show in the next sections that it is indeed the
most favorable case for squeezing. The effective cooperativity is then
related to the usual cooperativity parameter $C=g^{2}N/2\kappa \tau \gamma $
by 
\begin{equation}
\tilde{C}=C\left| \frac{\Omega _{1}}{\Delta }\right| ^{2}\frac{\gamma }{%
\gamma _{0}}  \label{Ctilde}
\end{equation}
The cooperativity parameter determines the efficiency of the non-linear
effect. If $C$ - or $\tilde{C}$ - and $I_{2}^{in}$ are large enough, the
intracavity field exhibits a bistable behavior. It has been shown that the
field quantum fluctuations are maximal around the lower turning point of the
bistability curve, allowing for the outgoing field to be squeezed. In Ref. 
\cite{Vernac}, it has been demonstrated that this point is also interesting
for atomic squeezing. Therefore, we will study the fluctuations in the
vicinity of this turning point.

\subsection{Linearization and diffusion matrix}

The linearized equations obtained from (\ref{S+}-\ref{A2}) are 
\begin{equation}
\frac{d\delta S_{+}}{dt}=-(\gamma _{0}-i\tilde{\delta})\delta S_{+}+2i\tilde{%
g}\langle S_{z}\rangle \delta A_{2}+2i\tilde{g}\langle A_{2}\rangle \delta
S_{z}+F_{S_{+}}  \label{deltaS+}
\end{equation}
\begin{equation}
\frac{d\delta S_{-}}{dt}=-(\gamma _{0}+i\tilde{\delta})\delta S_{-}-2i\tilde{%
g}\langle S_{z}\rangle \delta A_{2}^{\dagger }-2i\tilde{g}\langle
A_{2}\rangle ^{\ast }\delta S_{z}+F_{S_{-}}  \label{deltaS-}
\end{equation}
\begin{equation}
\frac{d\delta S_{z}}{dt}=-\gamma _{0}\delta S_{z}-i\tilde{g}(\langle
S_{-}\rangle \delta A_{2}-\langle S_{+}\rangle \delta A_{2}^{\dagger
}-\langle A_{2}\rangle ^{\ast }\delta S_{+}+\langle A_{2}\rangle \delta
S_{-})+F_{S_{z}}  \label{deltaSz}
\end{equation}
\begin{equation}
\frac{d\delta A_{2}}{dt}=-(\kappa +i\Delta _{c})\delta A_{2}+i\frac{\tilde{g}%
}{\tau }\delta S_{+}+\sqrt{\frac{2\kappa }{\tau }}\delta A_{2}^{in}
\label{deltaA2}
\end{equation}
We can write Eqs. (\ref{deltaS+}-\ref{deltaA2}) and the hermitian conjugate
of (\ref{deltaA2}) in matrix form as 
\begin{equation}
\frac{d\ \left| \delta \xi (t)\right] }{dt}=-\left[ B\right] \left| \delta
\xi (t)\right] +\left| F_{\xi }\right]
\end{equation}
where

\begin{equation}
\left| \delta \xi (t)\right] =\left[ \delta A_{2}(t),\delta A_{2}^{\dagger
}(t),\delta S_{+}(t),\delta S_{-}(t),\delta S_{z}(t)\right| ^{T},
\end{equation}
\[
\left[ B\right] =\left( 
\begin{array}{ccccc}
(\kappa +i\Delta _{c}) & 0 & -i\tilde{g}/\tau & 0 & 0 \\ 
0 & (\kappa -i\Delta _{c}) & 0 & i\tilde{g}/\tau & 0 \\ 
-2i\tilde{g}\langle S_{z}\rangle & 0 & (\gamma _{0}-i\tilde{\delta}) & 0 & 
-2i\tilde{g}\langle A_{2}\rangle \\ 
0 & 2i\tilde{g}\langle S_{z}\rangle & 0 & (\gamma _{0}+i\tilde{\delta}) & 2i%
\tilde{g}\langle A_{2}\rangle ^{\ast } \\ 
i\tilde{g}\langle S_{-}\rangle & -i\tilde{g}\langle S_{+}\rangle & -i\tilde{g%
}\langle A_{2}\rangle ^{\ast } & i\tilde{g}\langle A_{2}\rangle & \gamma _{0}
\end{array}
\right) 
\]
and 
\begin{equation}
\left| F_{\xi }(t)\right] =\left[ \sqrt{2\kappa /\tau }\delta A_{2}^{in}(t),%
\sqrt{2\kappa /\tau }\delta A_{2}^{in\dagger
}(t),F_{S_{+}}(t),F_{S_{-}}(t),F_{S_{z}}(t)\right| ^{T}
\end{equation}

The covariance matrix $\left[ G(t)\right] $ is defined by 
\begin{equation}
\left[ G(t)\right] =\left| \delta \xi (t)\right] \left[ \delta \xi (0)\right|
\end{equation}
and the diffusion matrix by 
\begin{equation}
\left\langle \left| F_{\xi }(t)\right] \left[ F_{\xi }(t^{\prime })\right|
\right\rangle =\left[ D\right] \text{ }\delta (t-t^{\prime })  \label{D}
\end{equation}
The values of the atomic diffusion coefficients can be derived from the
quantum regression theorem or operator identities. The complete diffusion
matrix is given in the Appendix. The variances of the spin components and
their correlation functions are the elements of the zero time correlation
matrix $\left[ G(0)\right] $, which satisfies \cite{Carmichael} 
\begin{equation}
\left[ B\right] \left[ G(0)\right] +\left[ G(0)\right] \left[ B\right]
^{\dagger }=\left[ D\right]  \label{G0}
\end{equation}
Inverting Eq. (\ref{G0}), one obtains $\left[ G(0)\right] ,$ and,
consequently, the spin variances. We proceed with the calculation of the
minimal variance as in the first Section.

\section{Self Spin Squeezing}

\subsection{Minimal variance calculation\label{resultselfsqueezing}}

We first study the case of a coherent input field $A_{2}^{in}$. We have
studied a wide range of parameters. First, in Fig. 2, we have plotted the
minimum variance $\Delta S_{\min }$ as a function of the cooperativity $%
\tilde{C}$ for a typical steady state point ($\tilde{\delta}=10$, $\delta
_{c}=0$, $I_{2}=25.2$).

We see that the squeezing naturally increases with the cooperativity, the
non-linear interaction increasing, but saturates a little under 30 \% when $%
\tilde{C}$ exceeds 100. As it is a relatively accessible value for
experiments, we will often choose this value for $\tilde{C}$ in the
following.

Then, for a fixed cooperativity $\tilde{C}$, we have optimized, for each
value of $\tilde{\delta}$, the values of $\delta _{c}$ and $I_{2}$ yielding
the best atomic squeezing, that is the lowest $\Delta S_{\min }$. The
results are reported in Table I\ for $\tilde{C}=100$.

\bigskip

\begin{tabular}{|l||l|l|l|l|l|}
\hline
$\tilde{\delta}$ & 0 & 5 & 10 & 15 & 20 \\ \hline
$\delta _{c}$ & 0 & -0.2 & 0 & -0.4 & -0.2 \\ \hline
$I_{2}$ & 0.25 & 6.5 & 25.2 & 56.5 & 100 \\ \hline
$\Delta S_{\min }$ & 0.713 & 0.716 & 0.72 & 0.72 & 0.728 \\ \hline
\end{tabular}
\newline

{\it Table I. Minimal spin variances for the effective two-level system as a
function of $\tilde{\delta}$ (\~{C} is equal to 100, the other parameters
are adjusted for optimal noise reduction).}

We see that the squeezing does not vary much with $\tilde{\delta},$ and is
always close to 28-29\%. Increasing the cooperativity does not improve the
squeezing much: one saturates at a squeezing value of almost 30\%. The fact
that the minimum variance is rather independent from the triplet $(\tilde{%
\delta},\delta _{c},I_{2})$ when the cooperativity increases shows that
there exists an optimal turning point for the system that reaches a steady
state independent from the above mentioned triplet as soon as the
cooperativity is large enough. Furthermore, the minimal variance seems to be
rather indifferent to the atomic noise for large values of $\tilde{C}$. The
following sections aim to explain these {\it a priori} surprising results.

\subsection{Outgoing field spectrum}

As mentioned above, with the existence of atomic squeezing, we can expect
squeezing for the outgoing field $A_{2}^{out}.$ To analyze field squeezing,
a good variable is the noise spectrum in the frequency domain obtained from
the variance matrix $\left[ V_{out}\left( \omega \right) \right] $ defined
by 
\begin{equation}
\langle \left| \delta A_{2}^{out}\left( \omega \right) \right] \left[ \delta
A_{2}^{out}\left( \omega ^{\prime }\right) \right| \rangle =2\pi \delta
\left( \omega -\omega ^{\prime }\right) \left[ V_{out}\left( \omega \right) %
\right] 
\end{equation}
where 
\begin{equation}
\left| \delta A_{2}^{out}\left( \omega \right) \right] =\left| 
\begin{array}{c}
\delta A_{2}^{out}\left( \omega \right)  \\ 
\delta A_{2}^{out\dagger }\left( \omega \right) 
\end{array}
\right] 
\end{equation}
is the vector Fourier transform of $\left| \delta A_{2}^{out}\left( t\right) 
\right] .$ The minimal and maximal component of the noise spectrum are then 
\begin{equation}
S_{out\ \min }\left( \omega \right) =\left[ V_{out}\left( \omega \right) 
\right] _{1,1}+\left[ V_{out}\left( \omega \right) \right] _{2,2}-2\left| 
\left[ V_{out}\left( \omega \right) \right] \right| _{1,2}
\end{equation}
\begin{equation}
S_{out\ \max }\left( \omega \right) =\left[ V_{out}\left( \omega \right) 
\right] _{1,1}+\left[ V_{out}\left( \omega \right) \right] _{2,2}+2\left| 
\left[ V_{out}\left( \omega \right) \right] \right| _{1,2}
\end{equation}
We have plotted in Fig. 3 these two components for a given set of parameters
of Table I. As expected, the field is squeezed (maximum squeezing around 12
\%), but the squeezing occurs over a frequency range having width $100\gamma
_{0}$, much larger than the value of order $\gamma _{0}$ one might have
expected from Eq. (\ref{deltaS+}-\ref{deltaA2}) where the atomic time
constant is $1/\gamma _{0}$. This is a consequence of the field-atom
interaction in the ''bad cavity limit'', as will be explained in the next
section.

\subsection{Adiabatic elimination of the intracavity field}

Unlike the three-level system, the effective two-level system we consider is
in\ what is called the ''bad cavity limit'': $\kappa \gg \gamma _{0}$, the
field evolves much faster than the atoms. We can therefore adiabatically
eliminate the field operators in the linearized Eqs. (\ref{deltaS+}-\ref
{deltaA2}) . One obtains a new set of equations 
\begin{equation}
\frac{d\left| \delta S\right] }{dt}=-\gamma _{0}\left[ B_{3}\right] \left|
\delta S\right] +\left| F_{S}^{\prime }\right]  \label{evolutionadiabatique}
\end{equation}
with $\left| \delta S\right] =\left[ \delta S_{+},\delta S_{-},\delta
S_{z}\right| ^{T}$, $\left| F_{S}^{\prime }\right] =\left[ F_{S_{+}}^{\prime
},F_{S_{-}}^{\prime },F_{S_{z}}^{\prime }\right| ^{T}$, 
\begin{equation}
\left[ B_{3}\right] =\left( 
\begin{array}{ccc}
1-i\bar{\delta}+4\tilde{C}\frac{s_{z}}{1+i\delta _{c}} & 0 & -2i\beta _{2}
\\ 
0 & 1+i\bar{\delta}+4\tilde{C}\frac{s_{z}}{1-i\delta _{c}} & 2i\beta _{2} \\ 
-i\beta _{2}-2\tilde{C}\frac{s_{-}}{1-i\delta _{c}} & i\beta _{2}-2\tilde{C}%
\frac{s_{+}}{1+i\delta _{c}} & 1
\end{array}
\right)  \label{B3}
\end{equation}
where 
\[
s_{i}=\left\langle S_{i}\right\rangle /N\quad (i=+,-,z) 
\]
In these equations there appears a new decay constant 
\[
\gamma ^{\prime }=\gamma _{0}\left[ 1+4\tilde{C}\frac{s_{z}}{1+\delta
_{c}^{2}}\right] 
\]
which is approximately equal to $\tilde{C}\gamma _{0}$ if $\tilde{C}\gg 1.$
If we take the values of the parameters of Fig. 3, we find $\gamma ^{\prime
}=100\gamma _{0}$ which corresponds exactly to the frequency bandwidth over
which the outgoing field is squeezed. The atomic spectrum is broadened as a
result of the interaction with the field. The new diffusion matrix defined
by $\langle \left| F^{\prime }(t)\right] \left[ F^{\prime }(t^{\prime
})\right| \rangle =\left[ D^{\prime }\right] \delta (t-t^{\prime })$ is
given by 
\begin{equation}
\left[ D^{\prime }\right] =\gamma _{0}N\left( \left[ D_{at}\right] +\tilde{C}%
\left( 
\begin{array}{ccc}
4s_{z}^{2} & 0 & -2s_{z}s_{+} \\ 
0 & 0 & 0 \\ 
-2s_{z}s_{-} & 0 & s_{+}s_{-}
\end{array}
\right) \right)  \label{dprime}
\end{equation}
where $\left[ D_{at}\right] $ is the atomic diffusion matrix calculated from
the quantum regression theorem (see Appendix).

Equation (\ref{dprime}) shows that $\left[ D^{\prime }\right] $ is the sum
of the usual atomic diffusion matrix $\gamma _{0}ND_{at}$, which is
proportional to the atomic decay rate $\gamma _{0}$ and the number of atoms $%
N$, and {\em another} matrix, which represents the contribution of the field
fluctuations, proportional to $N^{2}$, since $\tilde{C}\propto N$. For large
values of the cooperativity, the main contribution to the noise originates
from the field. This conclusion is valid only if 
\begin{equation}
\gamma _{0}\ll \tilde{C}\gamma _{0}\ll \kappa ;  \label{adiab}
\end{equation}
if $\tilde{C}$ becomes too large, the atoms evolve as fast as the field, and
the adiabatic elimination is no longer justified. When inequalities (\ref
{adiab}) hold, one is able to perform analytical calculations for the
minimum variance, considering only the larger terms in (\ref{dprime}) and (%
\ref{B3}), and using an optimal bistable point as in Table I. It is possible
to show that the lowest minimum variance $\Delta S_{\min }$ that can be
obtained is $1/\sqrt{2}=0.707$, in good agreement with the results announced
in Sec. \ref{resultselfsqueezing}. The optimal \ bistable point is obtained
considering the most favorable steady state for squeezing. Since the
squeezing originates from atom correlations we expect it to be maximum when
the modulus of coherence $S_{+}$ is a maximum. From the steady state
equations (\ref{statS+}-\ref{statS-}-\ref{statSz}), it is straightforward to
show that the maximum value of $\left| S_{+}\right| ^{2}$ is $N^{2}/16$,
when $4I_{2}=1+\bar{\delta}^{2}.$ In this case, $S_{z}=N/4$ and, integrating
(\ref{evolutionadiabatique}), one finds, for $\tilde{C}\gg 1,$%
\begin{equation}
\Delta S_{\min }=\frac{1}{\sqrt{2}}+O\left( \frac{1}{\tilde{C}}\right) 
\end{equation}
We see that the self squeezing limit in this system is approximately 30 \%,
as was found in the numerical simulations.

\subsection{Noise contributions to the atomic variance}

To confirm our statement that the squeezing can be traced to fluctuations in
the cavity field, we have computed the different contributions to the atomic
noise. Indeed, instead of calculating directly the atomic variances from the
time domain equations, it is possible to derive the atomic noise spectra in
the Fourier domain and then integrate the various contributions over the
frequency to get the variances. The linear response theory allows for
calculating the response of the atoms to the field excitation. The
fluctuations of the incident field cause the intracavity field to fluctuate,
which induces fluctuations for the atomic variables. These fluctuations are
coupled back to the field by polarization fluctuations, hence the non-linear
coupling between the intracavity field $\left| \delta A_{2}\right] $ and the
spin fluctuations $\left| \delta S\right] $. To these sources of
fluctuations, we must add the coupling with the vacuum surrounding modes of
the cavity and the noise due to the decay and pumping of the sublevels.
Taking all these fluctuations into account,\ the atomic variance matrix $%
\left[ V_{at}\left( \omega \right) \right] $ is the sum of four
contributions \cite{Vernac}: 
\begin{equation}
\left[ V_{at}\left( \omega \right) \right] =\left[ V_{f}\left( \omega
\right) \right] +\left[ V_{v}\left( \omega \right) \right] +\left[
V_{dip}\left( \omega \right) \right] +\left[ V_{int}\left( \omega \right) %
\right]
\end{equation}
in which

\begin{enumerate}
\item  $\left[ V_{f}\left( \omega \right) \right] $ represents the
contribution of the incident field whose fluctuations are modified by the
atoms

\item  $\left[ V_{v}\left( \omega \right) \right] $ is the contribution of
the vacuum fluctuations

\item  $\left[ V_{dip}\left( \omega \right) \right] $ corresponds to the
fluorescence emitted in the cavity\ mode

\item  $\left[ V_{int}\left( \omega \right) \right] $ is the interference
term between the vacuum fluctuations and the dipole fluctuations.
\end{enumerate}

In Fig. 4, are plotted the various contributions to the minimal component of
the spin noise spectrum. It is clear that the atomic noise is due entirely
to the fluctuations of the incident field, whereas the sum of the
contributions from the other terms is negligible. We have integrated the
different spectra to yield the contributions to the variances $\Delta
S_{\min }$ and present the results in Table II.

\bigskip

\begin{tabular}{|c|c|c|c|c|c|}
\hline
$\Delta S_{f}$ & $\Delta S_{v}$ & $\Delta S_{dip}$ & $\Delta S_{int}$ & $%
\Delta S_{\min }$ & $\Delta S_{f}/\Delta S_{\min }$ \\ \hline
0.701 & 1.414 & 1.373 & -2.772 & 0.716 & 97.9 \% \\ \hline
\end{tabular}

{\it Table II. Contributions to the atomic variance (value of the
parameters: }$\tilde{C}=100${\it , }$\bar{\delta}=12${\it , }$\delta
_{c}=-0.2${\it , }$I_{2}=40${\it , }$\gamma _{0}=\gamma /1000${\it , }$%
\kappa =2\gamma =5.2MHz${\it )}

We see that the contribution of the field represents about 98 \% of the
total variance for the minimal component, whereas all the other
contributions amount to only 2\% of the global noise. Consequently, the $%
\Lambda $ scheme provides a non-linear regime in which the field imprints
its fluctuations on the atoms and in which all the other causes of noise are
negligible compared to the non-linear interaction. We can thus easily reach
the squeezing limit predicted in the previous section, for much smaller
values of the cooperativity (50-100). In \cite{Vernac3}, for a classical
ground-excited state two-level system, the self-squeezing limit was 50 \%,
but, in order to approach that limit, much larger values of the
cooperativity were required (1000 to 10000).

\section{Squeezing transfer}

The second configuration that we have studied is the case where the input
light is a broadband squeezed vacuum. In this case, $\langle
A_{2}^{in}\rangle =0$. Note that we are no longer in a non-linear regime as
in the previous section. We have shown that, for the 3-level Raman scheme
considered in this paper and for the optimized transfer conditions given in
Ref. \cite{Vernac2}, the squeezing transfer from the field to the atoms is
almost 100 \%, even though we are not in the strong coupling regime as in 
\cite{Vernac2}. The only approximation made is the same as in Ref. \cite
{Vernac2}: the intracavity intensity stays negligible and the steady state
mean spin is then $\langle S_{z}\rangle =N/2$ (''small angle\
approximation'' \cite{Wineland}). Of course, even though $\langle
A_{2}^{in}\rangle =0$, the average number of intracavity photons is not 0,
but $\sinh ^{2}r$ for a squeezed vacuum, where $r$ is a squeezing parameter.
If the average photon number is much less than $N$, however, one can set 
\begin{equation}
\langle S_{+}\rangle =\langle S_{-}\rangle =0,\quad \langle S_{z}\rangle =%
\frac{N}{2}.
\end{equation}
Under these assumptions, the calculation of the minimum variance can be done
analytically \cite{Vernac2}, and one can deduce the optimal transfer
condition 
\begin{equation}
\bar{\delta}=\delta _{c}=0.  \label{transferpoint}
\end{equation}
In this case, the minimal variance can be expressed as a function of the
effective cooperativity $\tilde{C}$, the ratio $\rho =\gamma _{0}/\kappa $
and the squeezing rate $e^{-2r}$ as 
\begin{equation}
\Delta S_{\min }=1-\frac{2\tilde{C}}{(1+\rho )(1+2\tilde{C})}(1-e^{-2r}).
\end{equation}
As we pointed out, we are not in a strong coupling regime, which corresponds
to $\rho \tilde{C}\gg 1$. In our case, $\tilde{C}\sim 100$ and $\rho =\gamma
_{0}/\kappa \sim 1/2000\ll \gamma /\kappa $ (for a ground-excited state
(g-e) system with $\kappa =2\gamma ,$ $\rho =1/2$ and $\rho C\gg 1)$. For
large values of $\tilde{C}$, the factor in front of $(1-e^{-2r})$ is close
to unity. If we define the transfer efficiency $\eta $ by 
\begin{equation}
\eta =\frac{1-\Delta S_{\min }}{1-e^{-2r}},
\end{equation}
we find, for large $\tilde{C}$,

\begin{equation}
\eta =\frac{1}{(1+\rho )\left( 1+\frac{1}{2\tilde{C}}\right) }\simeq \frac{1%
}{(1+\rho )}\left( 1-\frac{1}{2\tilde{C}}\right) .
\end{equation}
In particular, for infinite cooperativity, we find the same limit of maximal
squeezing as in \cite{Vernac2}, but in a regime for which $\gamma _{0}\ll
\kappa $ ($\rho \ll 1,$ $\eta \sim 1-\rho )$. In Fig. 5, we have plotted the
minimum variance versus the squeezing of the input field $R_{in}=$ $%
1-e^{-2r} $ , for $\tilde{C}=100$ and two different values of $\rho $
corresponding to the case of our effective two-level system ($\rho =1/2000$%
), and the case of an g-e system ($\rho =1/2$). In particular, the $\Lambda $
system enables one to reach much higher squeezing values than in a classical
g-e two-level scheme. To understand the influence of the cooperativity on
the transfer, the efficiency versus $\tilde{C}$ is plotted in Fig. 6 in the
two cases discussed previously. The first conclusion is that the transfer is
close to 100 \% in the regime considered ($\rho =1/2000$) and is not limited
by the vacuum noise as in a g-e transition \cite{Vernac2}. Second, the
efficiency quickly saturates with $\tilde{C}$; a cooperativity of 100
enables a transfer of 99.5 \% of the squeezing from the field to the atoms ($%
\rho =1/2000$). We find again that the transfer efficiency is excellent for
very reasonable values of the cooperativity ($\sim 100$), which is usually
the limiting experimental factor.

\section{Validity of the two-level model}

In this section, we discuss the validity conditions for our effective
two-level model and see how the model can be modified slightly to extend its
range of validity. In carrying out the adiabatic elimination that took us
from the full, three-level equations to the two-level equations, two basic
assumptions were made. First, it was assumed that optical pumping terms of
order 
\[
\Gamma _{p}=\gamma \left| \Omega _{1}/\Delta \right| ^{2}
\]
could be neglected. Second, it was assumed implicitly that the probe field
was sufficiently weak to neglect terms of order $\left| \Omega _{2}/\Omega
_{1}\right| ,$ where 
\[
\hat{\Omega}_{2}=gA_{2};\text{ \ \ \ \ \ }\Omega _{2}=\left\langle \hat{%
\Omega}_{2}\right\rangle 
\]

If one adiabatically eliminates both the excited state - ground state
coherence and the excited state population operators from Eqs. (\ref{pi1}-%
\ref{pr}), the resulting contributions to the time evolution of the ground
state operators of optical pumping, transit time decay, and incoherent
pumping is 
\begin{mathletters}
\label{3to2}
\begin{eqnarray}
\dot{S}_{z} &=&-\left( \gamma _{0}+\Gamma _{p}\right) S_{z}+N\Gamma
_{p}/2+(\Lambda _{2}-\Lambda _{1})/2 \\
\dot{S}_{+} &=&-\left( \gamma _{0}+\Gamma _{p}\right) S_{+}-\Gamma _{p}\hat{%
\Omega}_{2}/\Omega _{1}  \label{3to2c}
\end{eqnarray}
where terms of order $\Gamma _{p}\left| \Omega _{2}/\Omega _{1}\right| ^{2}$
have been neglected. Aside from the term of order $\Gamma _{p}\left| \Omega
_{2}/\Omega _{1}\right| ,$ which we neglect for the moment but return to
shortly, these equations have basically the same structure that we used for
our effective two-state model. If $\Gamma _{p}\ll \gamma _{0}$, the two
models coincide. For arbitrary ratios of $\Gamma _{p}/\gamma _{0}$, one can
still use an effective two-level model if the assignments 
\end{mathletters}
\begin{mathletters}
\label{assign}
\begin{eqnarray}
\gamma _{0}(\text{two-level)} &=&\gamma _{0}(\text{three-level)}+\Gamma _{p}
\label{assigna} \\
\Lambda _{1}(\text{two-level)} &=&\Lambda _{1}(\text{three-level)}
\label{assignb} \\
\Lambda _{2}(\text{two-level)} &=&\Lambda _{2}(\text{three-level)}+N\Gamma
_{p}  \label{assignc}
\end{eqnarray}
are made [recall that $\Lambda _{1}+\Lambda _{2}=N\gamma _{0}]$. To check
this hypothesis, we set $\gamma _{0}($three-level$)=\Gamma _{p}=\gamma /1000$%
, so that $\Gamma _{p}$\ is no longer negligible compared to $\gamma _{0}$,
and we compare the two- and three-level calculations for self-squeezing when 
$\Delta =100\gamma $, $\Omega _{1}=\sqrt{10}\gamma $. For these parameters, $%
C=\tilde{C}=100$ and $\left| \Omega _{2}/\Omega _{1}\right| $ is typically
of order 0.01 at the point of optimal squeezing. The results, displayed in
Fig. 7, are in good agreement for both models. We have generalized the
results obtained in Sec. \ref{resultselfsqueezing} for $\Gamma _{p}/\gamma
_{0}\ll 1$ to arbitrary ratios $\Gamma _{p}/\gamma _{0}$. We would like to
point out that the steady state values (\ref{statS+}), (\ref{statS-}), (\ref
{statSz}) are modified by the presence of $\Gamma _{p}$, and the relation $%
4I_{2}=(1+\Gamma _{p}/\gamma _{0})^{2}+\bar{\delta}^{2}$ must be satisfied
in order to have the same optimal value of $\left| S_{+}\right| =N/4$.

To achieve better agreement between the effective two-level and full
three-level calculations, it may also necessary to include the $\left|
\Omega _{2}/\Omega _{1}\right| $ term in Eq. (\ref{3to2c}). In the previous
cases, this term can be neglected, but, for closed systems [$\Gamma _{p}\gg
\gamma _{0}($three-level$)],$ the term linear in $\left| \Omega _{2}/\Omega
_{1}\right| $ in Eq. (\ref{3to2c}) can slightly increase the value of $%
\left| \left\langle S_{+}\right\rangle \right| $ above the maximum value of $%
N/4$ it has for open systems [$\Gamma _{p}\ll \gamma _{0}($three-level$)]$.
As such the value of $\Delta S_{\min }$ can be reduced below $1/\sqrt{2}$.
This feature is seen in Fig. 8, where inclusion of the linear term in a
corrected two-level model brings the results into good agreement with the
full three-level calculation. Squeezing values of 35\% can be reached in
such a regime. The corrected two-level model also includes a slight
correction arising from a term linear in $\left| \Omega _{2}/\Omega
_{1}\right| $ that we neglected in going from Eq. (\ref{evA2}) to (\ref{A2})
for the field evolution. When this linear term is included, Eq. (\ref{A2})
is modified as 
\end{mathletters}
\begin{equation}
\frac{dA_{2}}{dt}=-(\kappa +i\Delta _{c})\text{ }A_{2}+\frac{i\tilde{g}}{%
\tau }\left[ S_{+}+\left( \hat{\Omega}_{2}/\Omega _{1}\right) \Pi _{2}\right]
+\sqrt{\frac{2\kappa }{\tau }}A_{2}^{in}  \label{modfield}
\end{equation}

The simplified two-level model considered in the earlier sections provides a
good understanding of squeezing in a three-level system and allows one to
optimize the parameters in a very simple way. It can be brought into full
agreement with the three-level model when $\left| \Delta _{1,2}\right| \gg
\gamma ,\left| \Omega _{1}\right| $ and $\left| \Omega _{2}/\Omega
_{1}\right| \ll 1$, provided one uses the prescription (\ref{assign}) for
the decay and incoherent pumping and Eq. (\ref{modfield}) for the field
evolution. Finally, we would like to point out that, although the value of $%
\gamma _{0}$ itself is not critical, the ratio of the decay constants of the
coherence $S_{\pm }$ to the population difference $S_{z}$ is important,
since it determines the maximal coherence that can be obtained from the
system, hence the maximal squeezing. Changing this ratio from 1 (as in our
simple two-level system) to 1/2 (as in a classical e-g system) would enable
one to achieve a 50\% squeezing limit.

\section{Conclusion}

Using a quantum model for an ensemble of three-level atoms in a Raman ($%
\Lambda $) type configuration, we have derived the atomic spin fluctuation
spectra and variances and have shown rigorously the occurrence of spin
squeezing between the two ground state sublevels. Unlike configurations in
which the squeezing occurs between two excited states or between a ground
state and an excited state, the squeezing obtained here has a long lifetime,
does not require a very strong coupling and, as a consequence, presents
substantial advantages for\ experimental realization.

Spin squeezing may occur in two different regimes. In the first one, the
non-linearity of the atomic ensemble is exploited to squeeze the intracavity
field, which in turn imprints squeezing on the atomic ensemble, yielding
self-spin squeezing. In the second one, the atomic ensemble has a linear
behavior. It cannot create squeezing in the intracavity field. However, if
one of the incoming fields is squeezed, the atom-field coupling in the
cavity enables a transfer of squeezing of almost 100 \%.

\section{Acknowledgments}

This work is supported by the U. S. Office of Army Research under Grant No.
DAAD19-00-1-0412, by the National Science Foundation under Grant No.
PHY-0098016 and the FOCUS Center Grant, and by the Michigan Center for
Theoretical Physics.

\section{Appendix}

We give here the expression of the diffusion matrix appearing in (\ref{D})
for the two-level system. The matrix elements of the upper 2$\times $2
diagonal sub-block of $\left[ D\right] $ are the correlation functions of a
broadband squeezed field, equal to the one of a single mode field squeezed
by a factor $e^{-2r}$ \cite{Vernac2} 
\begin{equation}
<\delta A^{in}(t)\delta A^{in\dagger }(t)>=\cosh ^{2}(r)\text{ }\delta
(t-t^{\prime })
\end{equation}
\begin{equation}
<\delta A^{in}(t)\delta A^{in}(t)>=1/2\sinh (r)\text{ }e^{i\theta }\delta
(t-t^{\prime })
\end{equation}
\begin{equation}
<\delta A^{in\dagger }(t)\delta A^{in\dagger }(t)>=1/2\sinh (r)\text{ }%
e^{-i\theta }\delta (t-t^{\prime })
\end{equation}
\begin{equation}
<\delta A^{in\dagger }(t)\delta A^{in}(t)>=\sinh ^{2}(r)\text{ }\delta
(t-t^{\prime })
\end{equation}
leading to a field diffusion matrix

\begin{equation}
\left[ D_{f}\right] =\frac{2\kappa }{\tau }\left[ 
\begin{array}{cc}
\cosh ^{2}(r) & 1/2\sinh (r)\text{ }e^{-i\theta } \\ 
1/2\sinh (r)\text{ }e^{i\theta } & \sinh ^{2}(r)
\end{array}
\right]  \label{Df}
\end{equation}
The matrix elements of the lower 3$\times $3 diagonal sub-block of $\left[ D%
\right] $ are the correlation functions of the atomic noise operators
appearing in Eqs. (\ref{deltaS+})-(\ref{deltaSz}). They were evaluated with
the Einstein generalized relations \cite{Cohen}. and grouped in the atomic
diffusion matrix $\left[ D_{at}\right] $:

\begin{equation}
\left[ D_{at}\right] =N\gamma _{0}\vspace{0.6cm}\left[ 
\begin{array}{ccc}
1+\frac{\lambda _{1}-\lambda _{2}}{2}-s_{z} & 0 & (1+\lambda _{1}-\lambda
_{2})\frac{s_{-}}{2} \\ 
0 & 1-\frac{\lambda _{1}-\lambda _{2}}{2}+s_{z} & (-1+\lambda _{1}-\lambda
_{2})\frac{s_{+}}{2} \\ 
(1+\lambda _{1}-\lambda _{2})\frac{s_{+}}{2} & (-1+\lambda _{1}-\lambda _{2})%
\frac{s_{-}}{2} & \frac{1}{2}+(\lambda _{1}-\lambda _{2})s_{z}
\end{array}
\right]  \label{Dat}
\end{equation}
where $\lambda _{1,2}=\Lambda _{1,2}/N\gamma _{0}$ are the dimensionless
incoherent pumping terms for one atom and $s_{\pm ,z}=\left\langle S_{\pm
,z}\right\rangle /N$ are the steady state values for one atom. The other
elements of $\left[ D\right] $ are equal to zero since there is no
correlation between atomic and field fluctuations at the same time.
Therefore, 
\begin{equation}
\left[ D\right] =\left( 
\begin{array}{cc}
\left[ D_{f}\right] & 0 \\ 
0 & \left[ D_{at}\right]
\end{array}
\right)
\end{equation}
where $\left[ D_{at}\right] $ and $\left[ D_{f}\right] $ are defined above (%
\ref{Df})-(\ref{Dat}).\bigskip

Fig 1. Three-level system in a $\Lambda $ configuration.

Fig 2. Minimal variance versus effective cooperativity $\tilde{C}$ for a
typical steady state point ($\tilde{\delta}=10$, $\delta _{c}=0$, $%
I_{2}=25.2 $).

Fig 3. Minimal (below) and maximal (above) spectra for the outgoing field
for the same bistable point as in Fig. 1 (the frequency unit is $\gamma _{0}$%
).

Fig 4. Contributions to the atomic noise spectrum (black): the field
contribution (light grey) is predominant over the sum of the three other
contributions, (dark grey). The bistable parameters are $\tilde{C}=100${\it %
, }$\bar{\delta}=12${\it , }$\delta _{c}=-0.2${\it , }$I_{2}=40$ as in Table
II.

Fig 5. Minimal variance $\Delta S_{\min }$ as a function of the input field
squeezing $R_{in}$ for $\rho =1/2000$ [$\Lambda $ system] and $\rho =1/2$
[classical 2-level system].

Fig 6. Transfer efficiency $\eta $ versus effective cooperativity $\tilde{C}$
for the same systems as in Fig. 5.

Fig 7. Minimal variance versus effective detuning $\tilde{\delta}$ in the
case when the ratio $\Gamma _{p}/\gamma _{0}=1$, calculated using the
2-level model with prescription (\ref{assign}) and the full 3-level model.

Fig 8. Minimal variance versus effective detuning $\tilde{\delta}$ in the
case of a closed system [$\Gamma _{p}\gg \gamma _{0}$], calculated using the
corrected 2-level model and the 3-level model. The small field term in Eq. (%
\ref{3to2c}) allows for going below the $1/\sqrt{2}$ minimal variance limit
and improving the squeezing (up to 35\% in this case).

\end{document}